\documentstyle[12pt]{article}
\def\beq{\begin{equation}}
\def\eeq{\end{equation}}

\begin{document}
\begin{titlepage}
\begin{flushright}
NSF-ITP-93-145 \\
BUTP-93/27 \\
PUTP-1433 \\
November 1993
\end{flushright}
\vspace{0.2in}
\begin{center}
{\Large \bf The Goldberger -- Treiman Relation, $g_A$ and $g_{\pi NN}$ at
$T\neq 0$ } \\
\vspace{0.4in}
{\bf V.L. Eletsky$^{\dagger}$,} \\
{\em Institute for Theoretical Physics, University of California \\
Santa Barbara, CA 93106 \\
and \\
Institute for Theoretical Physics, Berne University  \\
Sidlerstrasse 5, CH-3012 Berne, Switzerland} \\
\vspace{0.2in}
{\bf and} \\
\vspace{0.2in}
{\bf Ian I. Kogan$^{\dagger}$} \\
{\em Physics Department, Jadwin Hall \\
 Princeton University, Princeton, NJ 08540} \\
\vspace{0.5in}
{\bf   Abstract  \\ }
\end{center}

The Goldberger-Treiman relation is shown to persist in the chiral limit
at finite temperatures
to order $O(T^2)$. The $T$ dependence of $g_A$ turns out to be
the same as for $F_{\pi}$,
$g_{A}(T)=g_{A}(0)(1-T^2/12F^2)$,
while $g_{\pi NN}$ is temperature independent to
this order. The baryon octet ${\cal D}$ and ${\cal F}$ couplings also behave as
$F_{\pi}$
if only pions are massless in the pseudoscalar meson octet.

\vskip1.0in
\hrule height .2pt width 3in
\noindent$^{\dagger}$On leave of absence from: {\em Institute of Theoretical
and
Experimental Physics, Moscow 117259, Russia}
\end{titlepage}

In the recent years there has been an increasing interest in the study of
QCD at finite temperatures, both below and above the anticipated phase
transitiom. At high temperatures QCD is believed to be in a deconfined phase
of quarks and gluons, while in the low temperature region the appropriate
degrees of freedom are provided by hadrons. At sufficiently low $T$ the
dominant role is played by pions, other hadrons being suppressed by factors
of $\exp(-m_{had}/T)$. Using the pion gas approximation and effective chiral
Lagrangians, a number of problems
was considered regarding the changes in the vacuum structure and
hadron properties under the introduction of a finite temperature. Among
these were $T$ dependences of hadron masses and non-perturbative
condensates (for a review, see e.g.\cite{l}). It turned out that there is no
$T$ dependence of hadron masses in the leading order in the
thermal pion gas density, i.e. in order $O(T^2)$ (the only exception is the
pion mass $T$ dependence)\cite{en,ls,ei}. This is in fact a very general
consequence of $PCAC$ and current algebra\cite{l25}. The matrix elements of
correlators of hadronic currents usually do get $O(T^2)$ contributions which
result in parity and isospin mixing\cite{dei,ei}.
The $T$ dependence of the relevant mixing
coefficients determines the $T$ dependence of residues, i.e. matrix elements
of a current between vacuum and hadronic states, such as
$<0|A_{\mu}^a|\pi ^{b}(q)>=iq_{\mu}F_{\pi}\delta^{ab}$, where $A_{\mu}$ is
the axial isovector current and $F_{\pi}\simeq 90$\, MeV.
An interesting result of the $O(T^2)$ calculations is that the famous
Gell-Mann--Oaks--Renner relation,
$F_{\pi}^2 m_{\pi}^2 = -2m_q <\bar{q}q>$, holds to this order\cite{gl}.

In this paper we consider another famous relation, the Goldberger--Treiman
relation
\begin{equation}
g_A m_N=F_{\pi}g_{\pi NN}
\label{gt}
\end{equation}
at finite temperatures in the massless pion gas approximation.

It is well known that this relation follows when the requirement of the
axial current conservation is applied to the matrix element of the axial
current between nucleons,
\begin{equation}
<N(p)|A_{\mu}^a |N(p^{\prime})>=
\bar{N}(p)\frac{\tau^a}{2}\left(
g_A\gamma_{\mu}\gamma_{5}+h_{A}q_{\mu}\gamma_5\right)
N(p^{\prime}),~~~q=p-p^{\prime}
\label{me}
\end{equation}
The formfactor $g_A$ is finite at $q^2=0$ due to the pion pole in $h_A (q^2)$
which dominates in Eq.(\ref{me}) at $q^2\to 0$ (Fig.1). The constant
$g_{\pi NN}$ is defined by the phenomenological Lagrangian
${\cal L}^{\pi N}=ig_{\pi NN}\bar{N}\phi\gamma_{5}N$
($\phi\equiv\phi^{a}\tau^{a}$ and $\tau^a$ is a standard
isospin matrix).

The extension to finite $T$ is achieved by attaching a thermal pion loop to
the pole diagram in Fig.1 whereever possible. The relevant set of graphs is
presented in Fig.2 where a dash on pion loop indicates that this is a
pion from the heat bath. The first two diagrams correspond to the pion
residue and
wave function renormalization, while the last four ones involve $\pi N$
interactions.

A very convenient framework for considering correlation functions in
interacting pion gas is provided by the
method of effective chiral Lagrangians~\cite{w}. A particularly useful version
of
this method was developed by Gasser and Leutwyler~\cite{eff} by introducing
interactions of pions with external fields.
In this approach the generating functional of QCD
\begin{equation}
e^{iZ[v,a,s,p]}=\int [DA_{\mu}][Dq][D\bar{q}]
e^{i\int d^4 {\cal L} (q,\bar{q},G_{\mu\nu} ; v,a,s,p)}
\label{zq}
\end{equation}
is written in terms of effective meson theory
\begin{equation}
e^{iZ[v,a,s,p]}=\int [DU]e^{i\int d^4 {\cal L}_{eff}(U; v,a,s,p)}
\label{zu}
\end{equation}
where the $SU(2)_{R}\times SU(2)_{L}$ matrix $U$ contains the pion field,
$U(x)=\exp(i\phi /F)$.
Interaction with external vector, axial, scalar and pseudoscalar fields
$v_{\mu},a_{\mu},s,p$ is introduced through
\begin{equation}
{\cal L}={\cal L}^{0}+\bar{q}\gamma_{\mu}(v_{\mu}(x)+\gamma_{5}a_{\mu}(x))q-
\bar{q}(s(x)-i\gamma_{5}p(x))q
\label{L}
\end{equation}
where
\begin{equation}
{\cal L}^{0}=-\frac{1}{2g^2} G_{\mu\nu}^a G_{\mu\nu}^a+
\bar{q}\gamma_{\mu}(i\partial_{\mu}+A_{\mu})q
\label{L0}
\end{equation}
is the Lagrangian of massless $QCD$.
The external fields $v_{\mu},a_{\mu},s,p$ are hermitean $2\times 2$ matrices
in the flavor space. The quark mass matrix
${\cal M} =diag(m_u,m_d)$ is included into $s(x)$.
The real world of course corresponds to $v_{\mu}=a_{\mu}=p=0$ and
$s(x)={\cal M}$.
However, the introduction of external fields $v_{\mu}, a_{\mu}, p$ is
extremely helpful for obtaining bosonized versions of the corresponding
quark currents.

The effective Lagrangian in Eq.(\ref{zu}) is written as a series
\begin{equation}
{\cal L}_{eff}={\cal L}_2 +{\cal L}_4 +...
\label{Leff}
\end{equation}
according to the number of derivatives and/or quark mass factors.
Transformation properties of the matrix $U$ and the external fields under
$SU(2)_{R}\times SU(2)_{L}$ rotations
\begin{eqnarray}
U^{\prime}&=&RUL^{\dagger} \nonumber \\
v_{\mu}^{\prime}+a_{\mu}^{\prime}&=&R(v_{\mu}+a_{\mu})R^{\dagger}+
iR\partial_{\mu}R^{\dagger} \nonumber \\
v_{\mu}^{\prime}-a_{\mu}^{\prime}&=&L(v_{\mu}-a_{\mu})L^{\dagger}+
iL\partial_{\mu}L^{\dagger} \nonumber \\
s^{\prime}+ip^{\prime}&=&R(s+ip)L^{\dagger}
\label{tr}
\end{eqnarray}
dictate the structure of the effective
Lagrangians ${\cal L}_n$ \cite{eff}.
The lowest order one is given by
\begin{equation}
{\cal L}_2 =\frac{1}{4} F^2 {\rm Tr}
\left(\nabla_{\mu}U^{\dagger}\nabla_{\mu}U
+\chi^{\dagger}U+\chi U^{\dagger}\right)
\label{L2}
\end{equation}
where $\chi =2B(s+ip)$. The constant $F$ is the pion decay constant in the
chiral limit, and the constant $B$ is related to the quark condensate
$<\bar{q}q>=-F^2 B$. The external vector and axial field enter through the
covariant derivative
\begin{equation}
\nabla_{\mu}U=\partial_{\mu}U -i(v_{\mu}+a_{\mu})U + iU(v_{\mu}-a_{\mu})
\label{nabla}
\end{equation}
Extracting from Eq.(\ref{L2}) terms linear in $v_{\mu}$ and $a_{\mu}$,
vector and axial currents are expressed through the matrix $U$,
\begin{eqnarray}
V_{\mu}^i &=&\frac{1}{4}F^2 \, {\rm Tr}\;\tau^i
\left( [U^{\dagger},\partial_{\mu}U]-[U,\partial_{\mu}U^{\dagger}]\right) \\
A_{\mu}^i &=&\frac{1}{4}F^2 \, {\rm Tr}\;\tau^i
\left( \{ U^{\dagger},\partial_{\mu}U\}+\{U,\partial_{\mu}U^{\dagger}\}\right)
\label{VAU}
\end{eqnarray}
or, expanding in powers of the pion field and picking out for simplicity the
$"0"$ component,
\begin{eqnarray}
V_{\mu}^0 &=&\phi^{+}\partial_{\mu}\phi^{-}-\phi^{-}\partial_{\mu}\phi^{+}+
O(\phi^4) \\
A_{\mu}^0 &=&-F\partial_{\mu}\phi^{0}
\left(1-\frac{4}{3F^2}\,\phi^{+}\phi^{-}\right)-
\frac{2}{3F^2}\,\phi^{0}\,\partial_{\mu}\left(\phi^{+}\phi^{-}\right)
+O(\phi^5)
\label{VA0}
\end{eqnarray}
{}From Eq.(\ref{VA0}) using the thermal pion propagator
\begin{equation}
\Delta_{T}(0)=\int\frac{d^4 k}{(2\pi )^4}\,2\pi\delta(k^2)\,
\frac{1}{\exp (|k_0|/T)-1}=\frac{1}{12}T^2
\end{equation}
we get
\begin{equation}
A_{2a}=-\frac{4}{3}\Delta_{T}(0)\, A_{0}=-\frac{1}{9}\frac{T^2}{F^2}\, A_{0}
\label{2a}
\end{equation}
where $A_{0}=g_{\pi NN}F\, \bar{N}\gamma_{5}\tau^{0} N\, q_{\mu}/2q^2$
is the pole
contribution in Fig.1 and $A_{2a}$, $A_{2b}$,... are the amplitudes
corresponding to the diagrams in Fig.2.

The pion wave function renormalization diagram in Fig.2b is easily estimated
using the interaction part of ${\cal L}_2$,
\begin{equation}
{\cal L}_{2}^{int}=\frac{1}{6F^2}
\left[(\vec{\phi}\partial_{\mu}\vec{\phi})^2-
\vec{\phi}^2(\partial_{\mu}\vec{\phi})^{2}\right]
\label{L2int}
\end{equation}
($\vec{\phi}^2\equiv\phi^{a}\phi^{a}$).
This interaction leads both to pion mass and wave function renormalization.
As shown in Ref.\cite{gl}, the pion mass is not renormalized in the
chiral limit,
$m_{\pi}(T)=m_{\pi}(0)(1+T^2/48F^2)$. The wave function renormalization
then gives
\begin{equation}
A_{2b}=\frac{2}{3}\Delta_{T}(0)\, A_{0}=
\frac{1}{18}\frac{T^2}{F^2}\, A_{0}
\label{2b}
\end{equation}
The diagrams of Fig.2a and 2b determine the $T$ dependence of $F$ obtained
in Ref.\cite{gl}
\begin{equation}
F^2 (T)=F^2 (0)\left[1+\left(-2\cdot \frac{1}{9}+\frac{1}{18}\right)
\frac{T^2}{F^2}\right]=F^2 (0)\left[1-\frac{1}{6}\frac{T^2}{F^2}\right]
\label{F}
\end{equation}
It should be noted that specific contributions of diagrams in Fig.2a and 2b
may change under the change of parameterization of the matrix $U$, however
the net result is invariant.

To calculate the remaining four diagrams in Fig.2 we use an elegant
extension of the effective chiral Lagrangian to $\pi N$ interactions described
in Ref.\cite{geo} and generalized to external fields in Ref.\cite{gss}.
The lowest order (in the number of derivatives) $\pi N$
Lagrangian is given by
\begin{equation}
{\cal L}^{\pi N}_{1}=
\bar{N}\left(i\gamma_{\mu}D_{\mu}-m_{N}+
\frac{i}{2}g_{A}\gamma_{\mu}\gamma_{5}u_{\mu}\right)N
\label{piN}
\end{equation}
where $D_{\mu}=\partial_{\mu}+\Gamma_{\mu}$. The chiral connection
$\Gamma_{\mu}$ and the axial-vector object $u_{\mu}$ are given by
\begin{eqnarray}
\Gamma_{\mu}&=&\frac{1}{2}
\left(u^{\dagger}\partial_{\mu}u+u\partial_{\mu}u^{\dagger}\right) \\
u_{\mu}&=&i(u^{\dagger}\nabla_{\mu}u+u\nabla_{\mu}u^{\dagger})
\label{gu}
\end{eqnarray}
Here $u^2 = U$ and the
contact interactions of a nucleon with even and odd numbers of pions
are given by $\Gamma_{\mu}$ and $u_{\mu}$, respectively.
In terms of the pion field
\begin{equation}
\frac{1}{2}u_{\mu}=-\frac{1}{2F}\partial_{\mu}\phi +
\frac{1}{12F^3}\vec{\phi}^2\partial_{\mu}\phi -
\frac{1}{12F^3}(\vec{\phi}\partial_{\mu}\vec{\phi})\phi
\label{umu}
\end{equation}
The last two terms in the above provide the correction due to the diagram
of Fig.2c,
\begin{equation}
A_{2c}=-\frac{1}{3}\Delta_{T}(0)\, A_{0}=-\frac{1}{36}\frac{T^2}{F^2}\,
A_{0}
\label{2c}
\end{equation}

The remaining diagrams in Fig.2 do not contribute to order $T^2$.
The diagram in Fig.2d gives a vanishing contribution because it contains
$\partial_{\mu}\Delta_{T}(x)$ at $x=0$.
The diagram in Fig.2e has two extra powers of
the thermal pion momentum and will hence be of order $T^4$.
The diagram in Fig.2f vanishes by the
equation of
motion, since the external nucleons are {\em on-shell}.
Notice that the  diagram of Fig.2e
is likely to generate a new tensor structure for the matrix element in
Eq.(\ref{me}) proportional to
$q_{\mu}-n_{\mu}(nq)/q^2$ (transversality maintained),
where $n_{\mu}$ is the 4-velocity vector of the heat bath.
This is similar to what happens~\cite{eek}
for two-point correlation functions at $T\neq 0$: to order $T^2$ they are
Lorentz covariant with  Lorentz covariance broken in order $T^4$.

Summing up contributions from diagrams of Fig.2 we get
\begin{equation}
A_{2}=-\frac{1}{12}\frac{T^2}{F^2}\, A_{0}
\label{2}
\end{equation}
Using conservation of axial current (new tensor structures to not appear in
order $T^2$) and taking into account that $m_N$ is not
shifted in order $T^2$~\cite{en,ls}, we are led to infer that
\begin{equation}
g_{A}(T)=g_{A}(0)\left(1-\frac{1}{12}\frac{T^2}{F^2}\right) \; ,
\label{ga}
\end{equation}
i.e. $g_A$ goes with temperature just as $F$.
{}From here we conclude that $g_{\pi NN}$ is $T$-independent to order $T^2$.

One can see all this at the level of diagrams. The diagram of Fig.2a together
with a half of the diagram of Fig.2b gives the $T$-dependence of $F$
(see Eq.(\ref{F}). The second half of the diagram in Fig.2b together with
the diagram in Fig.2c should then give the $T$-dependence of $g_{\pi NN}$
\begin{equation}
\frac{1}{2} A_{2b}+A_{2c}= \left(\frac{1}{36}-\frac{1}{36}\right)
\frac{T^2}{F^2}\, A_{0}=0
\label{g}
\end{equation}

We have extended this analisys to the case of $SU(3)_{L}\times SU(3)_{R}$ and
octet baryons in which case the meson-baryon chiral Lagrangian
\begin{equation}
{\cal L}_{MB}  =  {\rm Tr}\,
\left( i\bar{B}\gamma_{\mu}D_{\mu}B-m\bar{B}B+
\frac{1}{2}{\cal D}\,\bar{B}\gamma_{\mu}\{ u_{\mu},B\} +
\frac{1}{2}{\cal F}\,\bar{B}\gamma_{\mu}\gamma_{5}[u_{\mu},B]\right)
\label{mb}
\end{equation}
($B$ is the standard $SU(3)$ matrix including all octet baryons) involves
two couplings, $\cal D$ and $\cal F$, such that $g_A ={\cal D}+{\cal F}$.
Considering only pions in
the meson sector to be massless we get
\begin{equation}
\frac{{\cal D}(T)}{{\cal D}(0)}=\frac{{\cal F}(T)}{{\cal
F}(0)}=\frac{g_{A}(T)}{g_{A}(0)}=
\frac{F_{\pi}(T)}{F_{\pi}(0)}=1-\frac{1}{12}\frac{T^2}{F_{\pi}^2}
\label{last}
\end{equation}

The result that pion-nucleon coupling stays unchanged to first order in the
pion density is an amusing one. We should note however, that a number of
earlier model calculations which used a chiral mean field theory~\cite{rob},
a QCD sum rules type of approach~\cite{ind}, and topological chiral soliton
model for the nucleon~\cite{mb} suggested that this
dependence is rather mild up to
temperatures in the region of phase transition.
We feel that it is worth asking a question whether
other hadronic couplings involving pions share this feature.
We would expect, for example, the $\omega\rho\pi$ coupling to have a very
weak $T$ dependence, since by vector dominance it is related to the
$\pi\gamma\gamma$ coupling, and the axial anomaly is not renormalized by
temperature effects\cite{mu}. We plan to address this issue in future work.

This work was started and completed at ITP, Santa Barbara, during the
program "Strong Interactions at Finite Temperatures". We are thankful for
the hospitality extended to us at ITP.
We acknowledge J. Gasser and S. Treiman for useful conversations.
This  work was supported in part by Scweizerischer Nationalfonds and
the U.S. National Science Foundation under Grant No. PHY89-04035.

\pagebreak
\begin{center}
{\large\bf Figure Captions}
\end{center}
\begin{itemize}
\item
Fig. 1: Pion pole contribution to the matrix element of the axial current
over nucleon. Solid, dashed and wavy lines correspond to the nucleon, pion, and
the axial current.
\item
Fig. 2: Corrections to the pion pole contribution to first order in the density
of the thermal pion gas. A dash on a pion line denotes the thermal pion.
\end{itemize}

\newpage
\begin{thebibliography} {99}
\bibitem{l}    H. Leutwyler, in {\em Effective field theories of the
               standard model}, ed. U.-G. Mei{\ss}ner, World Scienific,
               Singapore, 1992.
\bibitem{en}   V.L. Eletsky, Phys. Lett. {\bf 245B} (1990) 229.
\bibitem{ls}   H. Leutwyler and A.V. Smilga, Nucl. Phys. {\bf B 342} (1990)
302.
\bibitem{ei}   V.L. Eletsky and B.L. Ioffe, Phys.Rev. {\bf D 47} (1993) 3083.
\bibitem{l25}  H. Leutwyler, {\em On the foundations of chiral perturbation
               theory}, preprint BUTP-93/24.
\bibitem{dei}  M. Dey, V.L. Eletsky, and B.L. Ioffe, Phys. Lett. {\bf 252B}
               (1990) 620.
\bibitem{gl}   J. Gasser and H. Leutwyler, Phys. Lett. {\bf 184B} (1987) 83.
\bibitem{w}    S. Weinberg, Physica {\bf 96A} (1979) 327.
\bibitem{eff}  J. Gasser and H. Leutwyler, Ann. Phys. (N.Y.) {\bf 158} (1984);
               Nucl. Phys. {\bf B 250} (1985) 465.
\bibitem{geo}  H. Georgi, "Weak Interactions and Modern Particle Physics",
               Benjamin/Cummings, Reading, MA, 1984.
\bibitem{gss}  J. Gasser, M.E. Sainio, and A. \v{S}varc, Nucl. Phys.
               {\bf B 307} (1988) 779.
\bibitem{eek}  V.L. Eletsky, P.J. Ellis, and J.I. Kapusta, Phys. Rev. {\bf D47}
               (1993) 4084.
\bibitem{rob}  R.D. Pisarski, Phys. Lett. {\bf 110B} (1982) 155.
\bibitem{ind}  J. Dey, M. Dey, and P. Ghose, Phys. Lett. {\bf B 165} (1985)
181.
\bibitem{mb}   V. Bernard and U.-G. Mei{\ss}ner, Ann. Phys. (N.Y.)
               {\bf 206} (1991) 50.
\bibitem{mu}   H. Itoyama and A.H. Mueller, Nucl. Phys. {\bf B 218} (1983) 349.

\end{thebibliography}
\end{document}